\documentclass[10pt,aps,reprint,amsfonts,amsmath,amssymb,twocolumn]{revtex4-1}

\usepackage{graphicx}
\usepackage[utf8]{inputenc}
\usepackage[breaklinks]{hyperref}
\usepackage{amsmath}
\usepackage{amssymb}
\usepackage{amsfonts}
\newcommand{\de}[2]{\frac{\text{d} {#1}}{\text{d}{#2}}}
\newcommand{\pd}[2]{\frac{\partial #1}{\partial #2}}

\newcommand{\kb}{k_\text{B}}
\newcommand{\td}{\tau_\text{drift}}
\newcommand{\tc}{\tau_\text{cross}}
\newcommand{\ignore}[1]{}
\begin{document}


\title{Interplay of fast and slow dynamics in rare transition pathways: the disk-to-slab transition in the 2d Ising model}


\author{Clemens Moritz}
\affiliation{Faculty of Physics, University of Vienna, Boltzmanngasse 5, 1090 Vienna, Austria}
\author{Andreas Tröster}
\affiliation{Institute of Materials Chemistry, Vienna University of Technology, Getreidemarkt 9, 1060 Vienna, Austria}
\author{Christoph Dellago}
\email[]{christoph.dellago@univie.ac.at}
\affiliation{Faculty of Physics, University of Vienna, Boltzmanngasse 5, 1090 Vienna, Austria}

\date{\today}

\begin{abstract}
Rare transitions between long-lived stable states are often analyzed in terms of free energy landscapes computed as functions of a few collective variables. Here, using transitions between geometric phases as example, we demonstrate that the effective dynamics of a system along these variables are an essential ingredient in the description of rare events and that the static perspective provided by the free energy alone may be misleading. In particular, we investigate the disk-to-slab transition in the two-dimensional Ising model starting with a calculation of a two-dimensional free energy landscape and the distribution of committor probabilities. While at first sight it appears that the committor is incompatible with the free energy, they can be reconciled with each other using a two-dimensional Smoluchowski equation that combines the free energy landscape with state dependent diffusion coefficients. These results illustrate that dynamical information is not only required to calculate rate constants but that neglecting dynamics may also lead to an inaccurate understanding of the mechanism of a given process.
\end{abstract}


\maketitle

\section{Introduction\label{sec:intro}}
Free energy landscapes are a ubiquitous tool in the study of rare events---such as nucleation events in phase transitions or conformational changes in biomolecules---as they are often used to get a first sense of which regions in configuration space are relevant for a given process. Calculating such a landscape requires the choice of a small number of order parameters that are supposed to capture the important degrees of freedom of the process in question. Examples of such order parameters include the potential energy, the size of the largest cluster in the system (in the case of nucleation), or quantities like the number of native contacts, radii of gyration or root-mean-square deviations from the target structure (in biological systems).

However, care has to be taken in the interpretation of such free energy landscapes, since the choice of coordinates is not unique, even if the relevant degrees of freedom are captured. In particular, any non-linear transformation of the coordinates will change the free energy landscape, including features like the existence and height of free energy barriers\cite{Frenkel2013}. Due to this arbitrariness, additional information on the dynamics along a chosen order parameter is required in order to reliably interpret a free energy landscape\cite{Chahine2007,Krivov2008,Hinczewski2010,Best2010a,Best2011a}. 

With higher-dimensional free energy landscapes additional complications arise, especially if there is no clear physical relationship between the different order parameters used, as is often the case. For instance, one may be tempted to associate the gradient in such a landscape with a probability flux. However, flow directions\cite{Trinkaus1983,Berezhkovskii2005a,Peters2009}, the channels in which the transition proceeds\cite{Northrup1983,Moro1998,Yang2007}, and isocommittor lines\cite{E.2006,E2005} are determined both, by the free energy landscape and the dynamics of the system along the different order parameters and may deviate significantly from naive expectations. In particular, the combination of a fast and a slow degree of freedom which we will encounter in this work, has been discussed using a toy model by \textcite{Metzner2006}.

\begin{figure}[btp]
  \begin{center}
    \includegraphics[width=\columnwidth]{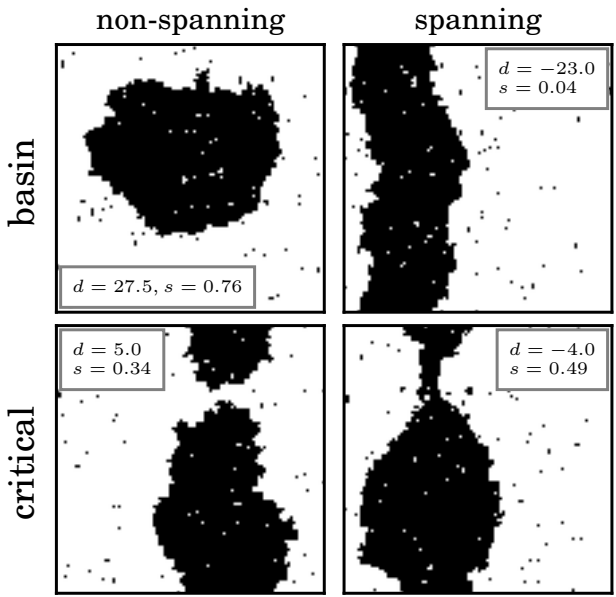}
  \end{center}
  \caption{Example configurations taken from simulations of the ferromagnetic Ising model at fixed magnetization of $m=0.36$ and temperature $T = 0.7 ~ T_\text{c}$. The configurations in the top row are taken from the stable basin of the disk- and the slab state respectively. The ones in the bottom row are critical configurations in the sense that their committor probabilities are close to $1/2$. $d$ and $s$ are order parameters discussed in section \ref{sec:model}.}
  \label{fig:configs}
\end{figure}

Often, especially in the context of rate calculations, one would like to project the system onto a single variable, called a reaction coordinate, that is not only capable of distinguishing between reactant and product state, but also describes the progress of the transformation in between. Due to the effects mentioned above, the free energy landscape may mislead us into a suboptimal choice for this reaction coordinate, which in turn negatively impacts the efficiency and accuracy of common rate calculation methods.

In the following, we use a simple model process to demonstrate some of the issues mentioned above: the disk-to-slab phase transition in the two-dimensional Ising model. Disk and slab phase are so called geometric phases\cite{Leung1990}, which arise due to periodic boundary conditions used in simulations. They are characterized by different cluster shapes---spherical, cylindrical and slablike in three dimensions---where the stability of the non-spherical shapes is due to the reduction of surface free energy that can be achieved by connecting a cluster to its periodic images. In two dimensions the number of geometries reduces to two: a disk- and a slab phase.

The disk-to-slab transition is an example of a transition between different cluster (or bubble) geometries that play a role in different physical situations~\cite{Troster2005,Singh2009a,Singh2011a,Santos2010,Kumar2011,Prestipino2015}, including the dewetting transition observed in volumes that are confined between hydrophobic surfaces~\cite{Lum1998,Nicolaides1989,Evans1999,Bolhuis2000,Leung2000,Luzar2000,Vishnyakov2003}. The mechanism of this dewetting has recently been discussed in detail by \textcite{Remsing2015}, who showed that it may involve the formation of a vapor bubble on one of the surfaces which subsequently changes its shape to a vapor tube that connects the plates. Under some conditions, this mechanism reduces the free energy barrier that has to be overcome, by, in essence, circumventing the states that involve a vapor tube size that is close to critical. The disk-to-slab phase transition can be seen as a simplified version of this process.

In addition, geometric phases pose an obstacle in the sampling of free energy landscapes~\cite{Neuhaus2003,Trebst2004} and they can be used to calculate surface tensions~\cite{Troster2011,Troster2012}.

The remainder of this paper is structured as follows: section \ref{sec:model} specifies the model that is investigated and introduces two order parameters that are then used to investigate the transition. In section \ref{sec:fe_committor} we discuss the free energy landscape and the distribution of committor probabilities as well as the apparent disconnect between the two. Section \ref{sec:calc_diff} explains how diffusion coefficients can be calculated reliably in this system, which are then used, in section \ref{sec:understanding}, to build a two-dimensional model of the system that predicts the distribution of committor probabilities. In section \ref{sec:discussion} we recapitulate our findings and discuss some of their implications.

\section{Model and Order Parameters\label{sec:model}}
We investigate the disk-to-slab phase transition in the two-dimensional Ising model with ferromagnetic nearest-neighbor interactions on a square lattice of size $N = 100 \times 100$ and a vanishing magnetic field. The Hamiltonian is given by 
\begin{equation}
    \mathcal{H} = \sum_{<ij,i'j'>} \sigma_{ij} \sigma_{i'j'},
\end{equation}
where the sum extends over all nearest-neighbor pairs and we have set the coupling constant to 1. The spins, denoted by $\sigma_{ij}$ with $i$ being the position in $x$-direction and $j$ the position in $y$-direction, can take on values of $+1$ (up) and $-1$ (down) and the magnetization per spin of the system is given by $m = (\sum_{i,j} \sigma_{ij}) / N$. Periodic boundary conditions apply in all directions. Since under these conditions the model is symmetric with respect to a flip of all spins, we restrict the discussion to positive $m$.

In this case, given a fixed temperature $T$ below the critical temperature $T_\text{c}$, three distinct phases can be found in order of descending magnetization: a homogeneous phase without a single dominant cluster of down spins, a disk phase where the bulk of the spins pointing down are in a single disk-shaped (\emph{non-spanning}) cluster and a slab phase where this cluster is in contact with its own periodic copies (\emph{spanning}). See Fig. \ref{fig:configs} for examples of disk- and slab-shaped clusters.

In this study, the dynamics of the system are given by a Metropolis Monte Carlo procedure with moves that keep the magnetization constant by only exchanging spin positions. We distinguish between local (Kawasaki~\cite{Kawasaki1966}) dynamics where only exchanges of neighboring spins are allowed and non-local dynamics that allow arbitrary exchanges. Local dynamics, which mimic mass transport in a more realistic system, are used in all simulations where dynamic properties are probed, while non-local moves are used for free energy calculations to enhance sampling. The unit of time $t$ is a Monte Carlo sweep, i.e., $N$ attempted single spin-exchange moves.

In the following discussion the reduced temperature is set to $\theta = T/T_\text{c} = 0.7$ and the magnetization to $m = 0.36$, where the slab state is metastable with respect to the disk state.

We will now introduce the order parameters $d$ and $s$, that we will use to describe transitions between the disk- and the slab state. For reference, Fig. \ref{fig:configs} shows examples of configurations together with their respective $d$- and $s$-values.

\subsection{The minimum distance/width order parameter $d$}
\begin{figure}[tbp]
  \begin{center}
    \includegraphics[width=\columnwidth]{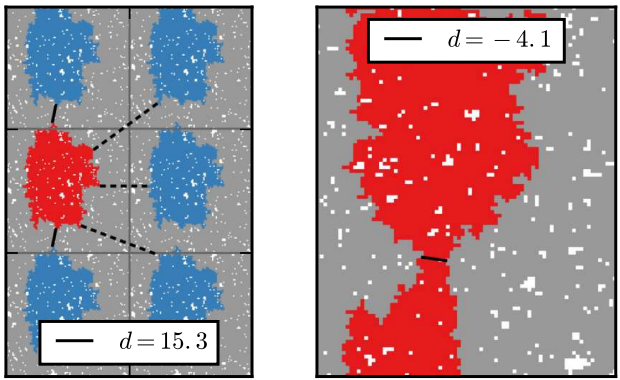}
  \end{center}
  \caption{Examples of the construction of the minimum distance/width order parameter $d$ for non-spanning clusters (left) and spanning clusters (right). The cluster indicated in red is the largest cluster of spins pointing down and the one in gray is the largest cluster pointing in the up direction. White spins do not belong to either of these clusters and the blue clusters are periodic copies of the red cluster. On the left, the smallest distance between the largest cluster and its periodic images is used as an order parameter while on the right the smallest distance between the two surfaces of the oppositely oriented cluster is measured. The latter is a proxy for the smallest width found along the length of the red cluster.}
  \label{fig:mindw}
\end{figure}

We first introduce an order parameter that is capable of distinguishing between disk- and slab-shaped clusters as well as between intermediate shapes. To do so, we identify the largest (geometric\cite{Schmitz2013,Binder2016}) cluster of spins, i.e. the largest connected cluster of spins that are neighbors of each other and point down. If this cluster is a non-spanning cluster, we characterize its shape by finding the shortest distance, $d_\text{image}$, between the surface of the cluster and the surfaces of any of its periodic images (see Fig. \ref{fig:mindw}). If the cluster is a spanning one, we instead characterize constrictions along its length by finding the smallest distance between the surfaces of the delimiting cluster of up spins, which is shown in gray on the right side of Fig \ref{fig:mindw}. This quantity is denoted by $d_\text{width}$.

The combined minimum distance/width coordinate $d$ is then given by
\begin{align}
  d = \begin{cases}
    d_\text{image} & \text{for non-spanning clusters} \\
    - d_\text{width} + 3 & \text{for spanning clusters}
  \end{cases},
\end{align}
where the number $3$ is added to -$d_\text{width}$ in order to close a gap in the possible values of $d$ that is due to the lowest possible $d_\text{image}$ and $d_\text{width}$ being $\sqrt{2}$ (when the closest spins of the two clusters are offset by one diagonal step) and $2$ (when there is a bridge of exactly one spin left), respectively. Consequently, for spanning clusters $d$ is less than or equal to $1$ and for non-spanning clusters $d$ is larger than $1$. Furthermore, the order parameter $d$ differentiates between different cluster shapes both for spanning and non-spanning clusters and yields information on the local structure of the cluster around the connecting bridge to its periodic images. 

The distance order parameter $d$ has units of one lattice constant.

\subsection{The shape order parameter $s$}
As we will see later, in order to arrive at a complete picture of the dynamics of the transition a second order parameter that characterizes the overall shape of the cluster is required. A straightforward way to do that for non-spanning clusters would be to use the ratio of the main moments of inertia of the cluster. However, this method breaks down as soon as the cluster is a spanning cluster since the main moments of inertia are no longer well defined.

\begin{figure}[tbp]
  \begin{center}
    \includegraphics[width=\columnwidth]{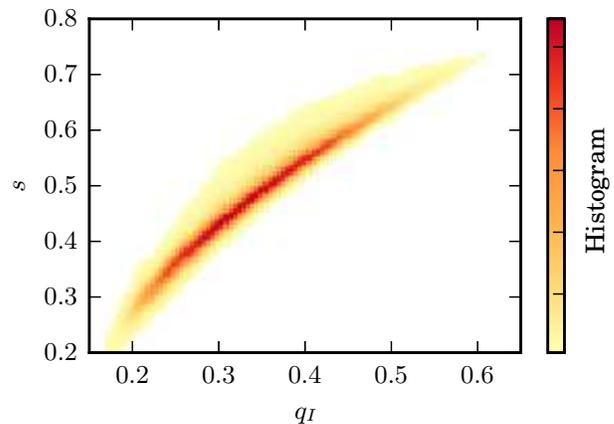}
  \end{center}
  \caption{Histogram of $q_I$-$s$ value pairs calculated from configurations that contain non-spanning clusters found in trajectories obtained from transition path sampling simulations~\cite{Dellago2002} of the disk-to-slab transition. $q_I$ is defined as the ratio of the smaller main moment of inertia of the cluster to the larger one. The $s$ values are highly correlated to the $q_I$ values indicating that they are an equally good measure for how disk-like a cluster's shape is.}
  \label{fig:corr_I_s}
\end{figure}

Instead, we will use a proxy in Fourier space, where, for a system with $n \times n$ spins, we first take a Fourier transform of the configuration yielding the components
\begin{equation}
  \tilde{\sigma}_{jk} = \frac{1}{n^2} \sum_{l,m=1}^{n} \sigma_{lm} \exp\left[ \frac{2 \pi i}{n} \left(jl + km\right) \right]
  \label{equ:fourier_trafo}
\end{equation}
and then take the (dimensionless) ratio
\begin{equation}
  s = \min \left( \frac{\left|\tilde{\sigma}_{01}\right|}{\left|\tilde{\sigma}_{10}\right|},\frac{\left|\tilde{\sigma}_{10}\right|}{\left|\tilde{\sigma}_{01}\right|} \right).
  \label{equ:f_def}
\end{equation}
Here, $\tilde{\sigma}_{10}$ and $\tilde{\sigma}_{01}$ are the longest wavelength components in $x$ and $y$ direction, respectively.

The order parameter $s$ takes on values from close to $0$---indicating configurations that are asymmetric with respect to an exchange of the $x$ and $y$ directions---to $1$ indicating that the largest cluster in the system is close to being symmetric. Its values are highly correlated to the ratio of the main moments of inertia for non-spanning clusters (see Fig. \ref{fig:corr_I_s}), but it is also well defined for spanning clusters.

\section{Free energies and the committor}\label{sec:fe_committor}
\begin{figure}[bp]
    \includegraphics[width=\columnwidth]{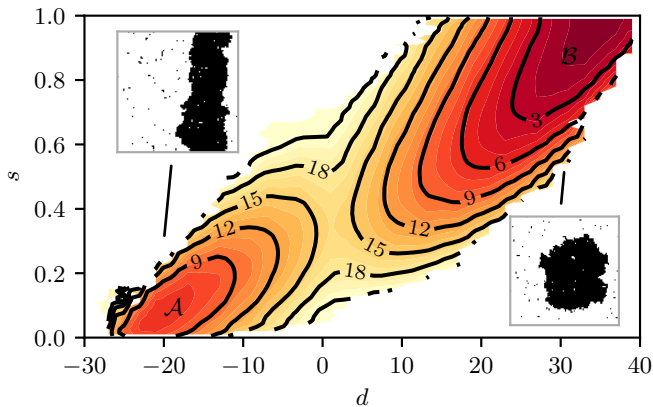}
    \caption{Free energy $F/k_\text{B} T$ (black isolines) of the 2D Ising model at magnetization per spin $m = 0.36$ and reduced temperature $\theta = 0.7$ as a function of the minimum distance/width order parameter $d$ and the shape order parameter $s$. The insets show examples of a slab-shaped~(top~left, basin $\mathcal{A}$) and a disk-shaped cluster~(bottom~right, basin $\mathcal{B}$). The zero-point of the free energy is chosen to coincide with the free energy minimum found in state $\mathcal{B}$. The data were obtained by a series of umbrella sampling simulations with harmonic biases and subsequent reweighting using WHAM~\cite{Ferrenberg1989,Kumar1992,Grossfield2013}.}
    \label{fig:plot_fe}
\end{figure}

We start our investigation of the mechanism of the disk-to-slab phase transition by computing the free energy landscape $\beta F(d,s) = -\log \left[P(d,s)\right]$, where $P(d,s)$ is the equilibrium probability density as a function of $d$ and $s$, $\beta = 1 / \kb T$, and $\kb$ is Boltzmann's constant. We do so by performing umbrella sampling simulations in the plane spanned by the $d$ and $s$ coordinates. Configurations are biased using a series of harmonic bias potentials that are centered at different points in the $d$-$s$ plane. The resulting probability distributions are then reweighted using the weighted histogram analysis method (WHAM)~\cite{Ferrenberg1989,Kumar1992}. Figure \ref{fig:plot_fe} shows the result of these calculations. The slab state, $\mathcal{A}$, is metastable with respect to the disk state, $\mathcal{B}$, by $6.4 ~ \kb T$ and the two states are separated by a barrier with a height of roughly $8~k_\text{B} T$ measured from the minimum within state $\mathcal{A}$.

\begin{figure}
    \begin{center}
        \includegraphics[width=\columnwidth]{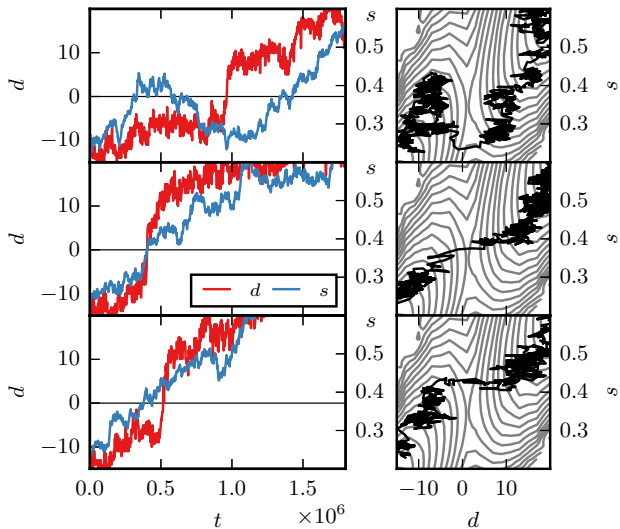}
    \end{center}
    \caption{Example trajectories of the disk-to-slab transition obtained from transition path sampling. The order parameters $d$ and $s$ are shown as a function of time (left) and as they appear projected onto the free energy landscape (right, isolines are placed $1 ~ k_\text{B} T$ apart). Time is given in units of Monte Carlo sweeps. Note that the transition over the ridge of the free energy landscape at $d=0$ progresses at a timescale that is much faster than the evolution of the $s$ coordinate.}
    \label{fig:trajs}
\end{figure}

In order to assess the quality of the two chosen order parameters $d$ and $s$, transition path sampling~\cite{Dellago2002} (TPS) simulations using aimless shooting~\cite{Peters2006a} and variable trajectory lengths have been used to obtain trajectories transitioning from state $\mathcal{A}$ to state $\mathcal{B}$. Configurations are assumed to have reached the basins if the value of $d$ is smaller than $-15$ or greater than $26$, respectively. The trajectories consist of segments with a length of $2000$ MC sweeps each and the shooting point is randomly shifted forward or backward in time by $20$ segments or the same shooting point as in the previous trajectory is chosen. Figure \ref{fig:trajs} shows example trajectories obtained from these simulations.

\begin{figure}[tbp]
    \includegraphics[width=\columnwidth]{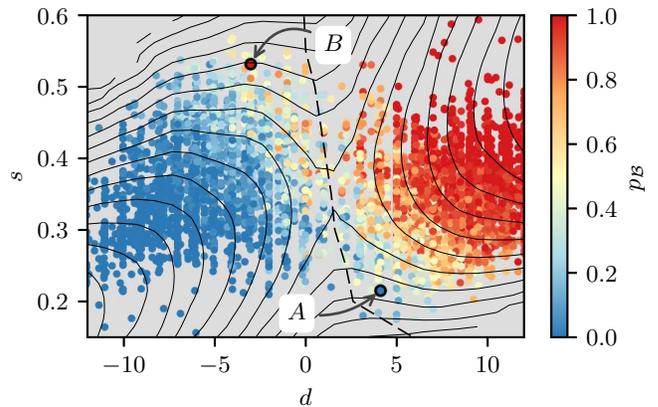}
    \caption{\label{fig:plot_commit}Committor probabilities $p_\mathcal{B}$ calculated for a section of the free energy landscape in the $d$-$s$ plane shown in Fig. \ref{fig:plot_fe}. The black lines are isolines of the free energy drawn in intervals of $k_\text{B} T$. The colored dots indicate the committor probabilities towards the disk state and the dashed black line represents the ridge of the free energy landscape that separates the basins of attraction of the disk and the slab state. The configurations marked with $A$ and $B$ are shown in Fig. \ref{fig:plot_commit_examples}.}
\end{figure}

As was mentioned in the introduction, a good reaction coordinate is able to characterize the progress of the transition. This means, that solely based on this coordinate one is able to predict the so called committor probability. The committor probability, $p_\mathcal{B}$, of a configuration is the probability of reaching state $\mathcal{B}$ before reaching state $\mathcal{A}$ averaged over trajectories started from the configuration. In our case, the averaging is done over trajectories computed with different random number generator seeds.

We computed committor values for configurations from pathways harvested in TPS runs. We estimated $p_\mathcal{B}$ by shooting trajectories from a given configuration and counting the number of times these trajectories reach state $\mathcal{B}$ first. The number of shots was chosen such that the statistical uncertainty of the estimated probability is smaller than $10 \%$~\cite{Dellago2002a}.

\begin{figure}[tbp]
    \includegraphics[width=\columnwidth]{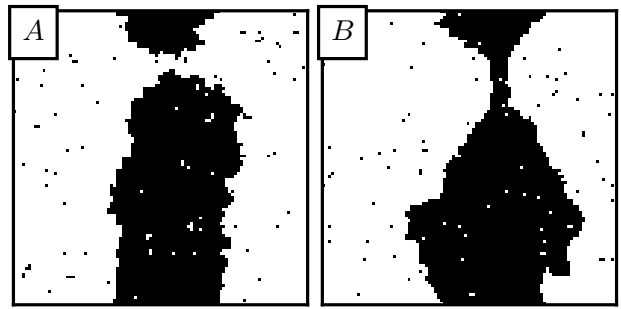}
    \caption{Example configurations $A$ and $B$ marked in Fig. \ref{fig:plot_commit}. These configurations are located at $(d=4.1,s=0.216)$ and $(d=-3.0,s=0.532)$ and have committor probabilities towards the disk state of $0.362$ and $0.958$ respectively.}
    \label{fig:plot_commit_examples}
\end{figure}

Figure \ref{fig:plot_commit} shows the resulting committor probabilities together with the free energy $\beta F(d,s)$. The shape of the free energy barrier suggests that the committor probability is determined by the value of the $d$-coordinate only. However, an examination of the committor probabilities calculated from simulation paints a different picture, in which there is a significant dependence of the committor on the value of the $s$-coordinate. In particular, some configurations that, based on the free energy landscape, seem to be in the basin of attraction of state $\mathcal{A}$ ($\mathcal{B}$) have a greater (smaller) than $1/2$ chance to evolve into the disk state $\mathcal{B}$. Such configurations, marked with $A$ and $B$, are shown in Fig. \ref{fig:plot_commit_examples}.

\begin{figure}[tbp]
    \includegraphics[width=\columnwidth]{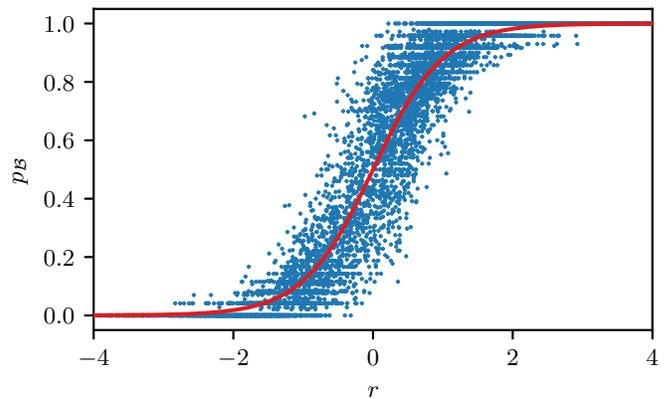}
    \caption{\label{fig:opt_committor}Committor probabilities towards the disk state as a function of a combined reaction coordinate $r = 0.18~d + 6.4~s - 2.6$. The parameters have been determined using a likelihood maximization procedure\cite{Peters2006a}. The red line represents the committor model function used for the optimization.}
\end{figure}

At this point one may suspect that we have not yet captured all the relevant degrees of freedom that determine the progress of the transition. However, Fig. \ref{fig:opt_committor}---which shows an optimized coordinate obtained using a likelihood maximization method due to \textcite{Peters2006a}---suggests that the committor can be fairly accurately described using only $d$ and $s$. In the following we will reconcile this apparent disconnect by including the dynamics along $d$ and $s$ into our considerations.

We start by noting that in the trajectories shown in Fig. \ref{fig:trajs}, the system spends a much larger amount of time close to the free energy barrier than it takes to cross the barrier to the other side. The $s$-coordinate evolves slowly because a large number of spins need to change collectively in order to change the shape parameter $s$ appreciably. Hence, for the dynamics chosen for the system, changing $s$ corresponds to the diffusive rearrangement of spins over long distances. In contrast, close to the barrier region, a change in the $d$-coordinate only requires the movement of relatively few spins and hence changes along the $d$-direction proceed much faster. As we will see later, this difference in relaxation times along the $d$- and $s$-direction strongly influence the distribution of committor probabilities. We will characterize this dynamical behavior by measuring the coordinate dependent diffusion coefficient tensor $D_{ij} (\vec{x})$, where $\vec{x} = (d,s)$.

\section{Characterizing System Dynamics}\label{sec:calc_diff}
In order to determine $D_{ij}$, we first employ the method of Im and Roux~\cite{Im2002,Pan2008} to calculate mean square displacements (MSDs) along the two coordinate directions that are corrected (up to first order) for the drift of the order parameters that is due to the underlying free energy landscape. Starting from equilibrium configurations within narrow ranges of $d$- and $s$-values, we generate short, unbiased trajectories. Using the coordinate $d$ as an example, the MSD is then estimated by
\begin{equation}\label{equ:msd-im-roux}
    \left<\delta \tilde{d}^2\right>_t = \left<\left(\tilde{d}(t) - \left< \tilde{d} \right>_t\right)^2\right>_t.
\end{equation}
Here $\tilde{d}(t) = d(t) - d(0)$ and the averages $\left<\ldots\right>_t$ are taken over the trajectories at time $t$ from their start. This procedure takes the mean drift (given by $\left< \tilde{d} \right>_t$) , as well as the slightly different starting points of the trajectories into account.
\begin{figure}[bp]
  \begin{center}
    \includegraphics[width=\columnwidth]{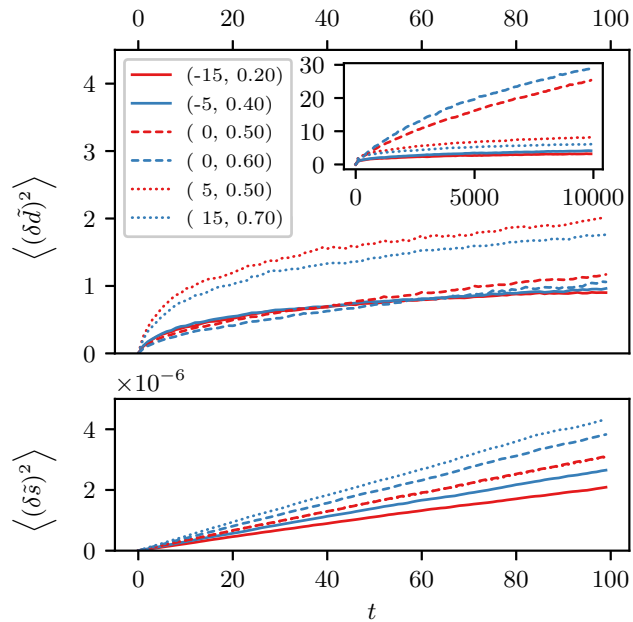}
  \end{center}
  \caption{Estimates for local mean square displacements measured using the method of Im and Roux~\cite{Im2002,Pan2008,Peters2009}. Different lines correspond to different starting points of the trajectories, which are given in the legend as $d$-$s$ pairs. Notice the distinct non-linear behavior along the $d$ direction (top), indicating the presence of strong memory effects at short timescales. The starting points at the top of the free energy barrier (dashed lines) show a markedly different behavior for long times (inset) due to the fact that two groups of trajectories separate---those developing towards the slab basin and those headed towards the disk basin.}
  \label{fig:msd-im-roux}
\end{figure}

Figure \ref{fig:msd-im-roux} shows the results of this calculation. In principle, linear fits to the data shown then yield the diffusion coefficients $D_{XX}$. While this is certainly true for the $D_{ss}$ component, the MSDs along the $d$-direction are highly non-linear even for long times, making it impossible to reliably determine the value for $D_{dd}$ this way.

\begin{figure}[tbp]
  \begin{center}
    \includegraphics[width=\columnwidth]{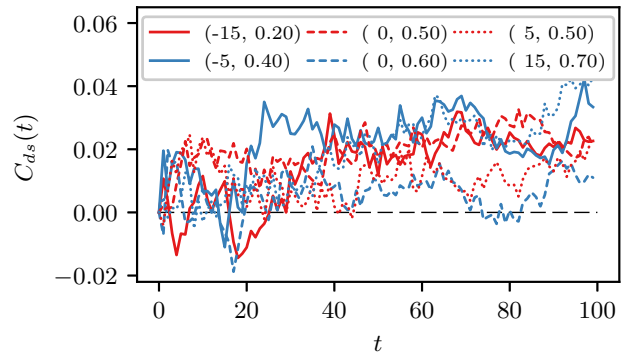}
  \end{center}
  \caption{Correlation $C_{ds}$ of mean displacements given by equation \eqref{equ:corr-im-roux}. The displacements in $d$- and $s$- direction are largely uncorrelated from each other.}
  \label{fig:corr-im-roux}
\end{figure}

\begin{figure}[tbp]
  \begin{center}
    \includegraphics[width=\columnwidth]{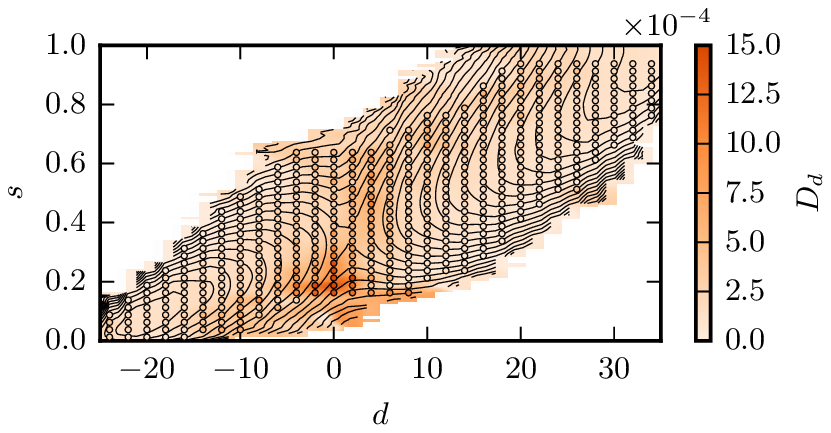}
    \includegraphics[width=\columnwidth]{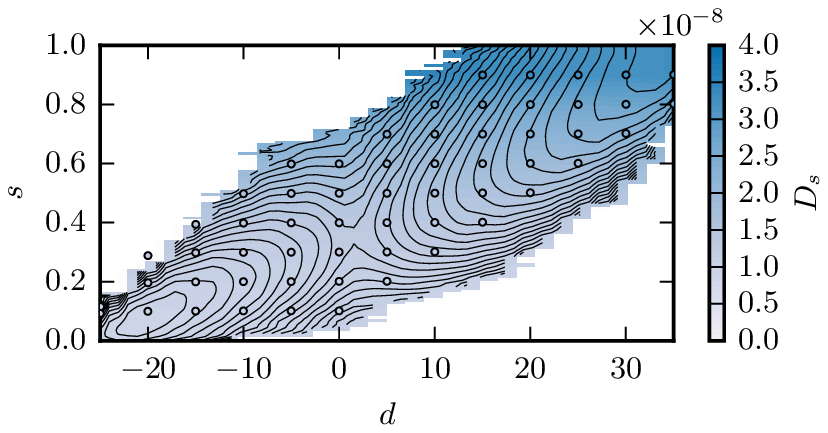}
  \end{center}
  \caption{Measured diffusion coefficients (circles) and cubic spline interpolation of these values (background) for the $D_d$ component (top) and the $D_s$ component (bottom). The diffusion coefficients are measured using a likelihood maximization method due to Hummer and coworkers~\cite{Hummer2005,Best2010a} in the $d$-direction and the mean-square displacement based ansatz by Im and Roux~\cite{Im2002,Pan2008,Peters2009} in the $s$-direction. The unit of time $t$ is 1 Monte Carlo sweep and the black lines indicate the free energy landscape where the lines are separated by $1~k_\text{B} T$.}
  \label{fig:diff_coeff}
\end{figure}

Before we address this problem, note that Fig. \ref{fig:corr-im-roux} shows the correlations
\begin{equation}\label{equ:corr-im-roux}
C_{ds}(t) = \frac{\left<(\delta \tilde{d})(\delta \tilde{s})\right>_t}{\sqrt{\left<(\delta \tilde{d})^2\right>_t}\sqrt{\left<(\delta \tilde{s})^2\vphantom{\delta \tilde{d})^2}\right>_t}},
\end{equation}
which are small and, hence, in the following we will treat the fluctuations in $d$- and $s$-direction to be independent of each other by assuming $D_{ij}$ to have the diagonal form
\begin{equation}
  \label{equ:diag_diff_tensor}
  \left(D_{ij}\right)\left(\vec{x}\right) = \begin{pmatrix}D_d\left(\vec{x}\right) & 0 \\ 0 & D_s\left(\vec{x}\right) \end{pmatrix}.
\end{equation}
The values for $D_s$ obtained by fitting to the MSDs are shown in the lower panel of Fig. \ref{fig:diff_coeff}.

In order to obtain reliable results for $D_d (\vec{x})$, we use a Bayesian method introduced by Best and Hummer~\cite{Hummer2005,Best2010a}, which is based on discretizing order parameter axes into bins followed by the maximization of the likelihood
\begin{equation}
    L = \prod_{\alpha} \left( e^{t_\alpha R} \right)_{i_\alpha,j_\alpha}.
\end{equation}
Here, $R$ is the matrix of transition rate constants between the bins, $t_\alpha$ is the lag time between two measurements of the order parameter and the product runs over all transitions of the system observed in a set of trajectories. The components of $R$ obey the conditions
\begin{align}\label{equ:rate_matrix}
  R_{ij} = \begin{cases} -\sum_{l \neq i} R_{il} & \text{if } i = j, \\
    R_{ji} P_i / P_j & \text{if } i < j,\end{cases}
\end{align}
where $P_i$ is the equilibrium probability of finding the system in bin $i$. Assuming that only the transition rates between neighboring bins are non-zero (i.e., we assume that the system transitioning from bin $i$ to $j$ has to pass through all bins in between), leaves us with $\mathcal{N}-1$ independent components, where $\mathcal{N}$ is the number of bins used. The diffusion coefficient between bin $i$ and $i+1$ along a coordinate $X$ with a bin width of $\Delta X$, is then given by~\cite{Hummer2005,Bicout1998}
\begin{equation}
  D_{i,i+1} = \Delta X^2 R_{i+1,i} \left( \frac{P_i}{P_{i+1}}\right)^{1/2}.
\end{equation}
The advantage of this method is that we can choose a lag time $\tau = t_\alpha$ that is long enough, such that the memory effects seen in the mean square displacements subside while we can still deal with the considerable drift that occurs within this lag time.

Since $D_d$ is considered to be independent of the $s$-direction and the evolution along the $s$-direction is slow, we can perform this calculation independently in parallel slices along lines of constant $s$. These slices have a width of $\Delta s = 0.025$ and each of these slices is split into bins of size $\Delta d = 2$. As suggested in Refs. \onlinecite{Hummer2005,Best2010a}, we count the number of transitions between these bins, $N_{ij}$, for a set of trajectories and subsequently vary the rate parameters $R_{ij}$ in order to maximize the log-likelihood
\begin{equation}
  \begin{aligned}
  \log \tilde{L} &= \sum_{i,j} N_{ij} \log\left[\exp\left(\tau R \right)_{ij}\right]\\
    & \qquad + \sum_i \frac{\left(D_{i+1,i} - D_{i,i-1}\right)^2}{2 \varepsilon^2},
  \end{aligned}
\end{equation}
where the second sum is a smoothing prior. We take the $P_i$ from previous free energy calculations. The matrix exponential $\exp\left(\tau R \right)$ is evaluated using a scaling-and-squaring method~\cite{Al-Mohy2009} as implemented in the python package scipy~\cite{Jonesa}. The value of $\varepsilon$ was chosen to be $3\cdot10^{-4}$, which is on the order of the eventual values obtained for $D_d$, in order to smooth large statistical fluctuations.

\begin{figure}[tbp]
  \begin{center}
    \includegraphics[width=\columnwidth]{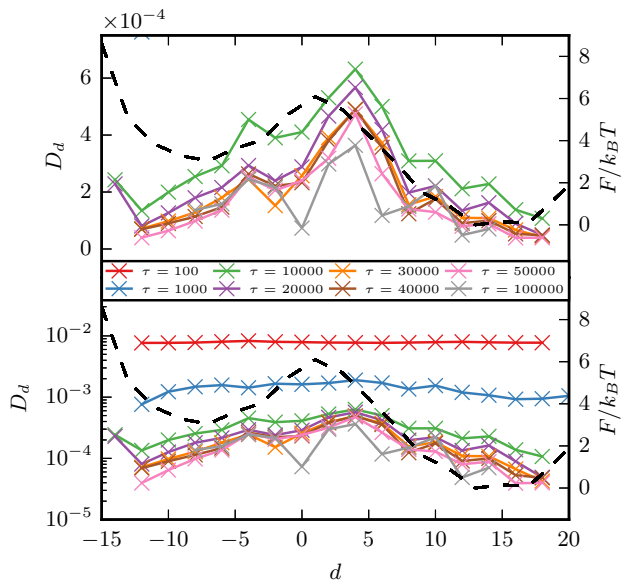}
  \end{center}
  \caption{Diffusion coefficient $D_{d}$ measured using the likelihood maximization method as a function of the chosen lag time $\tau$ for the $s = 0.375$-$0.400$ slice, chosen here as an example. Both panels show the same data but the lower one uses a logarithmic y-axis. The dashed, black line indicates the free energy profile along the slice. Time is measured in units of Monte Carlo sweeps.}
  \label{fig:lkh_max_lag_times}
\end{figure}

The lag time is chosen by varying the lag time until one observes a plateau in the resulting diffusion coefficients. Figure \ref{fig:lkh_max_lag_times} shows the results of such a calculation as a function of lag time for a single slice. At $\tau = 30\,000$ the values of the diffusion coefficients seem to have reached a plateau while statistics are still reasonably good. This lag time is used to calculate the values of $D_d$ shown in the upper panel of Fig. \ref{fig:diff_coeff}.

\section{Understanding the Committor}\label{sec:understanding}
In this section we will demonstrate that the shape of the committor distribution can be largely explained on the basis of the order parameters we have already introduced, $d$ and $s$, if one includes the dynamics of the system. Based on the free energy landscape and the diffusion coefficients we have obtained, we will first identify and estimate the timescales involved in the transition in order to gain an intuitive understanding, followed by an approach that uses a Smoluchowski equation to model the process.

\subsection{A qualitative explanation}\label{sec:qual_underst}
We start by noting that the drift towards the stable basins along the $s$-direction is slow compared to the relaxation time along the $d$-direction. Hence, in the following calculation we assume that a local equilibrium along slices of constant $s$ develops and consequently crossing the barrier becomes a one-dimensional problem associated with a timescale $\tau_\text{cross}(s)$, similar to a previous discussion of multicomponent nucleation by \textcite{Trinkaus1983}.

We then compare $\tau_\text{cross}$ to the timescale $\tau_\text{drift}$ that is associated with the diffusion along the $s$-direction; if $\tau_\text{cross} \gg \tau_\text{drift}$, the system is likely to stay on the same side of the barrier, whereas if they are comparable, one expects to see a finite probability of crossing to the other side of the barrier.

\begin{figure}[tbp]
  \begin{center}
      \includegraphics[width=\columnwidth]{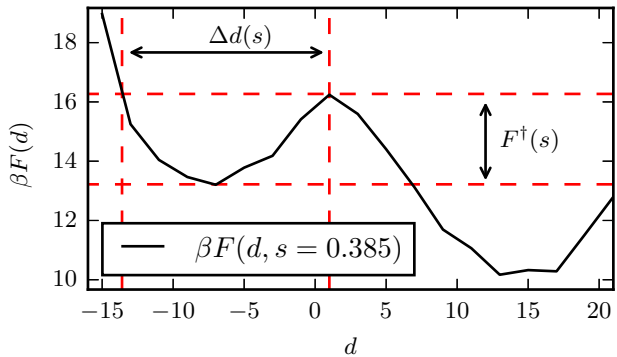}
  \end{center}
  \caption{Illustration of the quantities used to estimate the timescale of barrier crossing given by equation \eqref{equ:tau_cross}. The black line indicates the free energy $\beta F(d)$ along a slice of constant $s$. The quantities shown are used to estimate the crossing time from the basin of attraction of slab states towards the disk states. The calculation for crossing from disk to slab proceeds analogously using the width and depth of the disk basin (located at $d \approx 15$) instead.}
  \label{fig:explanation_timescales}
\end{figure}

Given the free energy landscape and the diffusion coefficient $D_d(s)$, we can estimate $\tc$ by
\begin{equation}
    \tc \sim \frac{(\Delta d(s))^2 }{D_d(s)} e^{\beta F^\dagger(s)}
  \label{equ:tau_cross}
\end{equation}
where $F^\dagger(s)$ is the height of the free energy barrier found when one looks only at a slice of the free energy landscape along a line of constant $s$ and $\Delta d$ is a length scale given by the width of the basin as seen along the slice (see Fig. \ref{fig:explanation_timescales}).

\begin{figure}[tbp]
  \begin{center}
    \includegraphics[width=\columnwidth]{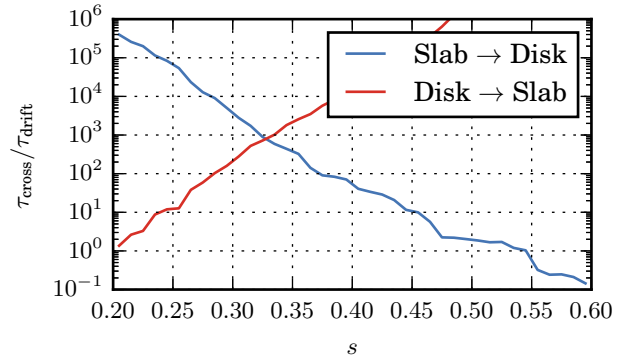}
  \end{center}
  \caption{Comparison of the timescales that are required to drift towards the stable basin, $\td$, and to cross the barrier to the other side of the transition, $\tc$, if one assumes equilibrium along a slice of constant $s$. The results have been obtained by numerical evaluation of Equ. \eqref{equ:timescale_fraction} using the free energy data shown in figure \ref{fig:plot_fe} and assuming $D_s / D_d = 3.2 \cdot 10^{-5}$, which is the value at the saddle point.}
  \label{fig:ratio_timescales}
\end{figure}

$\td$ is related to the drift of the system along the $s$-axis. As the system moves towards the stable basin and the barrier height $F^{\dagger}$ becomes larger, $\tau_\text{cross}$ grows exponentially as a function of $F^{\dagger}$. This change in barrier height can be used to estimate the time it takes until the system has moved to an area within the free energy landscape, where the rate of transitions has considerably changed relative to where the system has started from. We define a characteristic length $\Delta s$ as the distance at which the barrier height has increased by $\Delta F^\dagger = 1 ~ \kb T$. $\Delta s$ is then defined by setting
\begin{equation}
  1 = \beta \Delta F^\dagger(s) \approx \beta \pd{F^\dagger (s)}{s} \Delta s.
  \label{equ:delta_S_definition}
\end{equation}
One can now estimate $\td$ as the time it takes the system on average to undergo a change of $\Delta s$:
\begin{align}
  \td \sim \left|\frac{\Delta s}{\dot{s}}\right| \propto  \left|\frac{1 / \left(\beta \pd{F^\dagger (s)}{s}\right)}{D_s \beta \pd{F(s)}{s}}\right|
  \label{equ:tau_drift}
\end{align}
Here we have chosen $\dot{s}$ to be the mean velocity due to the drift caused by the slope in the free energy, i.e. $\dot{s} = - D_s \beta \pd{F(s)}{s}$. The ratio of the two timescales, given by
\begin{equation}
    \frac{\tc}{\td} = \frac{D_s}{D_d} \beta^2 (\Delta d (s))^2 \left|\pd{F(s)}{s} \pd{F^\dagger (s)}{s} \right| e^{\beta F^\dagger (s)},
  \label{equ:timescale_fraction}
\end{equation}
is shown in Fig. \ref{fig:ratio_timescales}.

One can see that in the outer regions with $s \lesssim 0.25$ and $s \gtrsim 0.45$ the crossing time becomes comparable to the drift time. Furthermore, if $\tc/\td$ is close to $1$ in one direction of the transition, it is orders of magnitude larger in the reverse direction. This suggests that in these regions the probability of crossing from one side of the barrier to the other is non-zero and, on the other hand, after the barrier has been crossed, the probability of going back is small. Hence, we arrive at the explanation of why the committor probabilities in these regions take on the values shown in Fig. \ref{fig:plot_commit}. It takes a long time to change the overall shape of the cluster and in this time fluctuations can occur that make the system cross the barrier. However, once the barrier has been crossed, the probability of going back the other way is small because the global shape of the cluster disfavors such fluctuations. In other words, there is a high probability that the bridge of spins that can be seen in the right panel of Fig. \ref{fig:plot_commit_examples} is severed before the cluster has changed its shape into a slab and, conversely, the likelihood of a bridge forming in the left panel before the cluster becomes disk-shaped is high.

\subsection{Smoluchowski analysis}\label{sec:smoluchowski}
A more accurate analysis of the system's behavior given the free energy landscape and the diffusion coefficients can be achieved by modeling the dynamics of the system using a two-dimensional Smoluchowski equation. A similar approach has previously been employed by \textcite{Metzner2006} in order to predict the behavior of a two-dimensional toy model. We use a generalized ansatz that takes into account state dependent diffusion coefficients\cite{Risken1984,Gardiner1985}. In this case, the probability density $\rho(\vec{x},t)$ satisfies the equation
\begin{equation}\label{equ:fp-smoluchowski}
  \begin{aligned}
    \pd{\rho}{t} &=
    - \sum_{i,j} \pd{}{x_i}\left[\left(\pd{D_{ij}}{x_j} - \beta D_{ij} \pd{F}{x_j}\right) \rho\right] \\
    & \qquad\qquad + \sum_{i,j} \frac{\partial^2}{\partial x_i \partial x_j}\left[\vphantom{\frac{\partial^2}{\partial x_i \partial x_j}} D_{ij} \rho\right],
  \end{aligned}
\end{equation}
where $\vec{x} = \left(d,s\right)$, $F = F(\vec{x})$ is the free energy shown before, and $D_{ij}(\vec{x})$ are position dependent diffusion coefficients. This form of the Fokker-Planck equation\footnote{The Smoluchowski equation \eqref{equ:fp-smoluchowski} is often rewritten in the form $\pd{\rho}{t} = \sum_i \pd{}{x_i} \left[\sum_j D_{ij}e^{-\beta F} \pd{}{x_j} \left( e^{\beta F} \rho \right)\right]$, that makes it immediately obvious that the equilibrium distribution is the stationary solution.} mimics a random walk on a potential of mean force given by the free energy $F$ and, at the same time, guarantees that the steady state solution is given by the equilibrium distribution $e^{-\beta F }$.

In this model the committor $p_\mathcal{B}(\vec{x})$ is then determined by the corresponding backward (or adjoint) equation~\cite{Gardiner1985,Metzner2006}
\begin{equation}\label{equ:back_kolmogorov}
  \begin{aligned}
      \sum_{i,j} \left(\pd{D_{ij}}{x_j} - \beta D_{ij} \pd{F}{x_j}\right) \pd{p_\mathcal{B}(\vec{x})}{x_i}\qquad&\\ + \sum_{i,j} D_{ij} \frac{\partial^2 p_\mathcal{B}(\vec{x})}{\partial x_i \partial x_j} &= 0.
  \end{aligned}
\end{equation}
The reflecting boundary condition
\begin{equation}
  \sum_i n_i D_{ij} \pd{p_\mathcal{B}\left(\vec{x}\right)}{x_j} = 0,
\end{equation}
where $\vec{n}$ is a normal vector on the domain, and the boundary conditions that surround regions $\mathcal{A}$ and $\mathcal{B}$
\begin{equation}
  \left. p_\mathcal{B}(\vec{x}) \right|_{\partial\mathcal{A}} = 0, \left. p_\mathcal{B}(\vec{x}) \right|_{\partial\mathcal{B}} = 1
  \label{equ:bound_conditions}
\end{equation}
close the set of equations.

\begin{figure}[tbp]
  \begin{center}
    \includegraphics[width=\columnwidth]{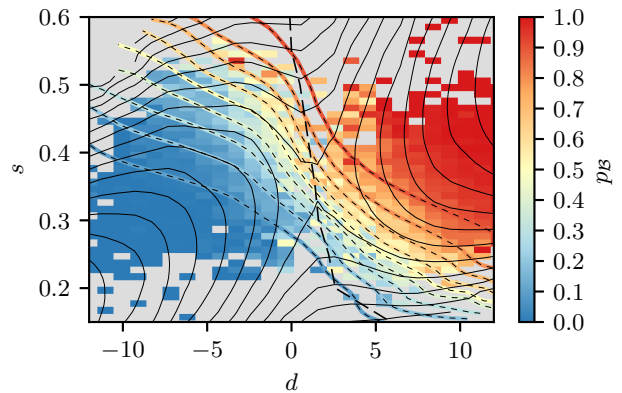}
  \end{center}
  \caption{Solution $q(d,s)$ of the adjoint Smoluchowski equation \eqref{equ:back_kolmogorov} (isolines with black dashes) using the free energy data obtained from simulation of the disk-slab transition and the diffusion coefficients shown in figure \ref{fig:diff_coeff}. The background color shows committor values shown in Fig. \ref{fig:plot_commit} averaged over squares of size $1.0 [d] \times 0.01 [s]$ and the dashed black line represents the ridge of the free energy landscape that separates the basins of attraction of the disk and the slab state.}
  \label{fig:committor_analytic}
\end{figure}

This system of equations can be solved using a finite difference scheme similar to the one employed in Ref. \onlinecite{Metzner2006} using the free energy and diffusion coefficients measured from simulation as input. The result of such a calculation is shown in Fig. \ref{fig:committor_analytic}. It recovers the basic features of the committor observed in the brute force simulations. In particular, the $p_\mathcal{B} = 1/2$ committor line significantly deviates from the ridge of the free energy landscape and the regions close to the barrier that have large or small values of $s$ have a significant probability of evolving to the other side of the free energy barrier.

\begin{figure}[tbp]
  \begin{center}
    \includegraphics[width=\columnwidth]{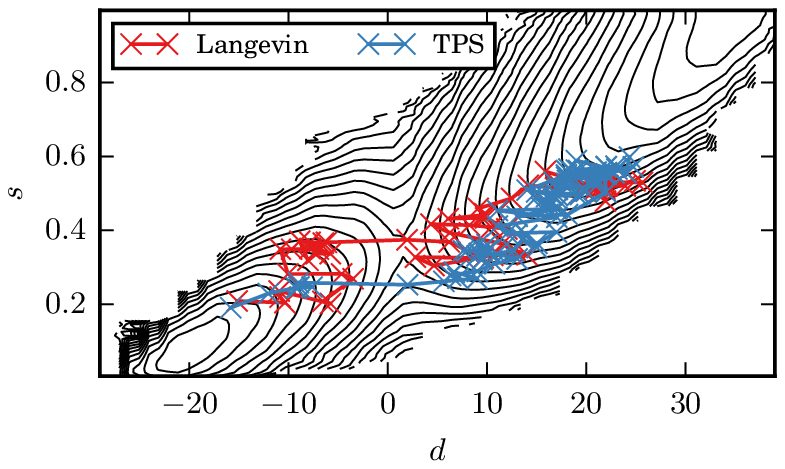}
    \includegraphics[width=\columnwidth]{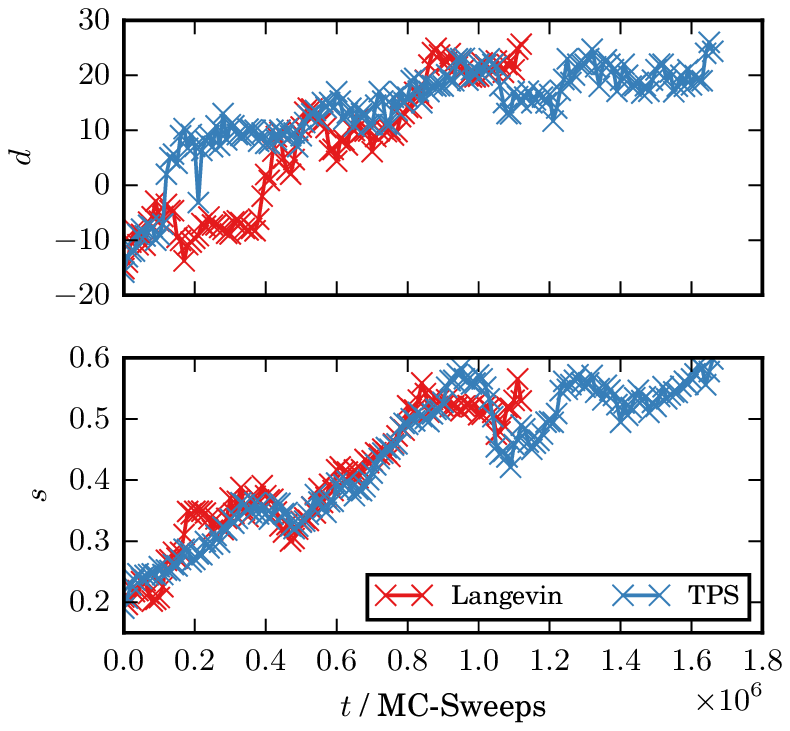}
  \end{center}
  \caption{Example trajectories obtained from TPS simulations (blue) and by integrating the Langevin equation \eqref{equ:langevin} (red) from a state at the saddle point as they appear projected onto the $d$-$s$-plane (top) and as a function of time (bottom).}
  \label{fig:langevin_trajectory}
\end{figure}

\begin{figure}[tbp]
  \begin{center}
    \includegraphics[width=\columnwidth]{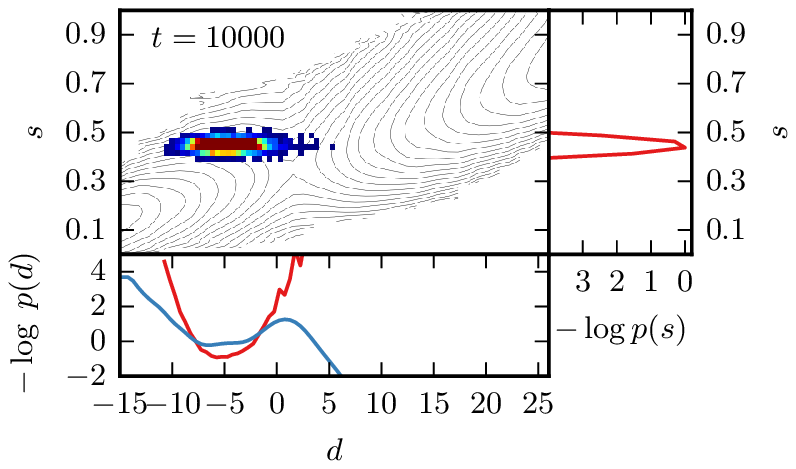}
    \includegraphics[width=\columnwidth]{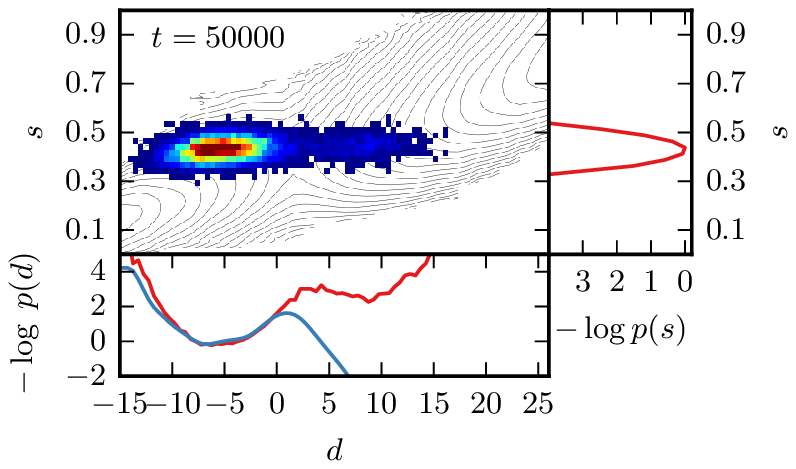}
    \includegraphics[width=\columnwidth]{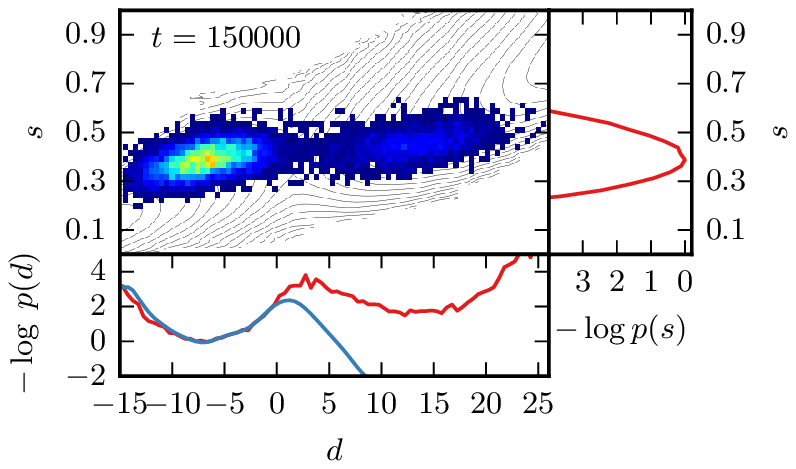}
    \vspace{-1cm}
  \end{center}
  \caption{Time evolution of the probability density function $\rho(\vec{x},t)$ as obtained by integrating trajectories using equation \eqref{equ:langevin} and subsequently averaging over the positions $x(t)$. The random walkers are started at position $d=-5$ and $s=0.45$. Red indicates high densities and blue low densities. In each of the plots, the bottom and the right plot show the negative logarithm of the probability density projected on each axis (red). In the bottom plot, the equilibrium free energy $\beta F(d,s=\left< s \right>_t)$ is shown (blue), where $\left< s \right>_t = \int \text{d}d ~ \int \text{d}s ~ s \rho(d,s;t)$ is the mean value of $s$ at the given time.}
  \label{fig:langevin_prob_distribution}
\end{figure}

We can also write down a corresponding Langevin-equation, whose realizations will---if evaluated using Ito's interpretation (i.e. by evaluating the drift and diffusion terms at the beginning of each step)---evolve according to Equ. \eqref{equ:fp-smoluchowski}\cite{Tupper2012}. It is given by
\begin{equation}\label{equ:langevin}
    \de{x_i(t)}{t} = \sum_j \left[\pd{D_{ij}}{x_j} - \beta D_{ij}\pd{F}{x_j}\right] + \sum_j g_{ij} \eta_j(t),
\end{equation}
where $\vec{x}(t)$ is the position of a random walker, $D_{ij}$, $F$ and their derivatives are evaluated at $\vec{x}(t)$, $\eta$ is delta-correlated white noise with zero mean and unity variance (i.e. $\left< \eta_i \right> = 0$, $\left<\eta_i(t) \eta_j(t')\right> = \delta (t-t') \delta_{ij}$), and $\sum_k g_{ik} g_{kj} = 2 D_{ij}$\cite{Risken1984}. The latter condition simplifies to $g_{ij} = \sqrt{2 D_{i}}\delta_{ij}$ if we assume the diffusion coefficient is diagonal and given by Equ. \eqref{equ:diag_diff_tensor}. Figure \ref{fig:langevin_trajectory} shows an example Langevin trajectory compared to a trajectory obtained by TPS and Fig. \ref{fig:langevin_prob_distribution} shows the probability distribution of finding a random walker as a function of time.

These two diagrams demonstrate two things: the trajectories generated using a full simulation and the ones generated from the Langevin equation very much look alike and, secondly, that indeed a relaxation towards equilibrium along the $d$ coordinate occurs before the system has time to evolve to the stable basins.

\section{Discussion}\label{sec:discussion}
We have investigated the disk-to-slab transition in the 2d-Ising model using a variety of methods, including a model based on a two-dimensional Smoluchowski equation. This model allows us to take the different timescales involved in this process into account: the fluctuations of the narrow bridge between the clusters (characterized by the $d$-coordinate) and the change of the overall shape of the cluster (captured by the $s$-coordinate). By comparing committor distributions obtained by brute-force using the full dynamics of the system, to the committor distribution obtained by solving the adjoint equation \eqref{equ:back_kolmogorov} (Fig. \ref{fig:committor_analytic}) we find that the dynamics of the transition are well described by the Smoluchowski equation \eqref{equ:fp-smoluchowski}.

This suggests that, with the coordinates $d$ and $s$, we have indeed captured the degrees of freedom that are most relevant for the disk-to-slab transition. Increased values of $s$, corresponding to disk-like clusters, enhance the likelihood of a trajectory started from a configuration to develop towards the disk state. While stated in this way, this result does not seem surprising, it is nonetheless interesting to note, that this effect only becomes apparent when one includes the dynamics of the system into consideration, while the free energy landscape alone does not capture this feature. In the disk-to-slab transition considered here, this is caused by the large difference in timescales associated with the crossing of the free energy barrier and the relaxation towards the stable basins that causes a local quasi-equilibrium to form where the global shape of the cluster is fairly stable, while its surface fluctuates.

Our analysis highlights the fact that, even in the context of a fairly simple model system, the dynamics of a system are crucial to identifying transition mechanisms, especially when one needs to use a combination of multiple order parameters to obtain a good reaction coordinate. In particular, including dynamics may help to reconcile free energy landscapes and committor probabilities, or their combination may point to the existence of other important degrees of freedom.

\begin{acknowledgments}
The authors thank Georg Menzl and Phillip L. Geissler for many enlightening discussions. C.M. is supported by an uni:docs fellowship of the University of Vienna. A.T. acknowledges support from the Austrian Science Fund (FWF) Project Nr. P27738-N28. The computational results presented have been achieved in part using the Vienna Scientific Cluster (VSC).
\end{acknowledgments}

\bibliography{paper_geomteric_phase_transitions}

\end{document}